\documentstyle[wrapft]{ptptex}
\notypesetlogo

\input epsf

\def\Tr{ {\rm Tr}}
\def\EXPECT#1{\langle #1 \rangle}
\def\KET#1{\vert{#1}\rangle}
\def\BRACKET#1#2{ \langle #1 \vert #2 \rangle}
\def\ORDER#1{${\cal O}\left( #1 \right)$}

\def\magweg#1{}

\title{Quantum Statistical Mechanics on a Quantum Computer}

\author{H. {\sc De Raedt}$^1$\footnote{E-mail: deraedt@phys.rug.nl}, 
        A.H. {\sc Hams}$^1$,
        K. {\sc Michielsen}$^2$, 
        S. {\sc Miyashita}$^3$ and 
        K. {\sc Saito}$^3$
}

\inst{$^{1}$Institute for Theoretical Physics and Materials Science Centre \\
      $^{2}$Laboratory for Biophysical Chemistry \\
      University of Groningen, Nijenborgh 4 \\
      NL-9747 AG Groningen, The Netherlands \\
\vskip 0.5cm 
      $^{3}$Department of Applied Physics, School of Engineering \\
      University of Tokyo, Bunkyo-ku, Tokyo 113  }

\recdate{
}

\abst{We describe a quantum algorithm to compute the density of states and
 thermal equilibrium properties of quantum many-body systems.
 We present results obtained by running this algorithm
 on a software implementation of a 21-qubit quantum computer
 for the case of an antiferromagnetic
 Heisenberg model on triangular lattices of different size.}
\begin{document}
\maketitle
\section{Introduction}
Recent theoretical work has shown that a Quantum Computer (QC)
has the potential of solving certain computationally hard
problems such as factoring integers
and searching databases much faster than a conventional computer.
\cite{SHORone,CHUANGone,KITAEV,GROVERzero,GROVERone,GROVERtwo}
The idea that a QC might be more powerful than an ordinary computer is based
on the notion that a quantum system can be in any superposition of states
and that interference of these states allows exponentially
many computations to be done in parallel.\cite{AHARONOVone}
This hypothetical power of a QC might be used to solve other
difficult problems as well, such as for example
the calculation of the physical properties of quantum many-body 
systems.\cite{CERFone,ZALKAone,TERHALone}
As a matter of fact, part of Feynman's original motivation
to consider QC's was that they might be used
as a vehicle to perform exact simulations of quantum mechanical 
phenomena.\cite{FEYNMAN}
For future applications it is clearly of interest to address the question how
to program a QC such that it performs a
simulation of specific physical systems.

In this paper we describe a quantum algorithm (QA) to compute
the equilibrium properties of quantum systems
by making a few statistically uncorrelated runs,
starting from random initial states.
Exploiting the intrinsic parallellism of the hypothetical QC
this QA executes in polynomial time.
En route this QA computes the density of states (DOS)
from which all the eigenvalues of the model Hamiltonian may be determined.
We test our QA on a software implemention of a 21-qubit QC
by explicit calculations for an antiferromagnetic
Heisenberg model on a triangular lattice.
\section{Theory}
Consider the problem of computing the physical properties of
a quantum system, described by a Hamiltonian $H$,
in thermal equilibrium at a temperature $T$.
In the canonical ensemble this equilibrium state
is characterised by the partition function $Z=\Tr \exp(-\beta H)$
where $\beta=1/k_B T$ and $k_B$ is Boltzmann's constant (we put $k_B=1$ and
$\hbar=1$ in the rest of this paper).
The dimension of the Hilbert space of physical states will be denoted by $D$.
If all the eigenvalues $\{E_i\,;\,i=1,\ldots,D\}$ of $H$
are known, we can make use of the fact
that $Z=\sum_{i=1}^D\exp(-\beta E_i)$
to reduce the computation of the physical properties to a
classical statistical mechanical problem which can be solved by
standard probabilistic methods.\cite{HAMMERSLEY}
For a non-trivial quantum many-body system the determination of
the eigenvalues is a difficult computational problem itself,
in practice as difficult as the calculation of $Z$.
In view of this we will from now on assume
that the eigenvalues of $H$ are not known.

The DOS
\begin{equation}
\label{epsilon}
{\cal D}(\epsilon)=\sum_i \delta(\epsilon-E_i) =
{1\over 2\pi}\int_{-\infty}^{+\infty} e^{it\epsilon}\, \Tr e^{-itH}\, dt
,
\end{equation}
determines the equilibrium state of the system. Indeed,
once ${\cal D}(\epsilon)$ is known $Z$ is easy to calculate:
\begin{equation}
 \label{integral}
Z= \int_{-\infty}^{+\infty} e^{-\beta \epsilon} {\cal D}(\epsilon)\, d\epsilon
.
\end{equation}
Integral~(\ref{integral}) exists whenever the spectrum of $H$ has a lowerbound, i.e.
${\cal D}(\epsilon)=0$ for $-\infty<\epsilon < \min_i E_i$.
Note that ${\cal D}(\epsilon)$ is a real-valued function.

With suitable (time-dependent) modifications of $H$,
the partition function plays the role of a generating function
from which all physical quantities of interest can be obtained.
Physical quantities such as the energy and specific heat are given by
\begin{equation}
 \label{min9}
E= \EXPECT{H} = {1\over Z}\int_{-\infty}^{+\infty}
\epsilon e^{-\beta\epsilon} {\cal D}(\epsilon)\, d\epsilon
,
\end{equation}
and
\begin{equation}
 \label{min8}
C= \beta^2(\EXPECT{H^2}-\EXPECT{H}^2) = \beta^2\left(
{1\over Z}
\int_{-\infty}^{+\infty} \epsilon^2 e^{-\beta\epsilon} {\cal D}(\epsilon)\, d\epsilon
- E^2\right)
,
\end{equation}
respectively. 
\magweg{Time-dependent properties, e.g. dynamic structure factors
can be obtained in this way as well.\cite{PEDROone} }

From~(\ref{epsilon}) it is clear that the computation of
${\cal D}(\epsilon)$ consists of two parts: 1) Compute the
trace of $e^{-itH}$ for many values of $t$ and
2) perform a (Fast) Fourier Transform of this data.
It is known how to carry out part 2 on a QC\cite{AHARONOVone,DIVINCENZOone,ECKERTone,CLEVEone}
so we focus on part 1.

The QA described below computes $e^{-itH}\KET{\psi}$
for any $\psi$ in ${\cal O} \left({\log D}\right)$ operations.
We now argue that in practice it will usually be sufficient
to determine a small fraction ($\approx {\cal O}\left( \log D\right)$) of the $D$
diagonal matrix elements of $e^{-itH}$.
Instead of computing diagonal matrix elements with respect to
a chosen complete set of basis states $\{\phi_n; n=1,\ldots,D\}$, we
generate random numbers $\{a_n; n=1,\ldots,D\}$ and construct the new state
\begin{equation}
\KET{\Phi}=\sum_{n=1}^D a_n \KET{\phi_n}
.
\end{equation}
The corresponding diagonal element of the time-evolution operator reads
\begin{equation}
\BRACKET{\Phi}{e^{-itH}\Phi}=
\sum_{n,m=1}^D a_n^* a_m^{\phantom{*}} \BRACKET{\phi_n}{e^{-itH}\phi_m}
.
\end{equation}
If we now generate the $a_i$'s such that
$\overline{a^*_n a_m}=\delta_{n,m}$
($\overline{x}$ denotes the average of $x$ over statistically independent realizations)
then
\begin{equation}
\overline{\BRACKET{\Phi}{e^{-itH}\Phi}}=
\sum_{n=1}^D \BRACKET{\phi_n}{e^{-itH}\phi_n}=
 \Tr e^{-itH}
.
\end{equation}
In other words, the trace of the time-evolution operator can be estimated
by random sampling of the states $\KET{\Phi}$.\cite{ALBEN}

A QC computes $\KET{e^{-itH}\Phi}$ just as easily as $\KET{e^{-itH}\phi_m}$
and in practice there would be no need to have a random state
generator: Switching the QC off and on will put the QC in some random initial
state.
However to compute $\BRACKET{\Phi}{e^{-itH}\Phi}$ we would have to store
the initial state ($\KET{\Phi}$) in the QC and project
$\KET{e^{-itH}\Phi}$ onto it, a rather complicated procedure.
Instead it is more effective to apply to $\KET{\phi_1}$ a random sequence
of Controlled-NOT operations to construct e.g. an entangled
random state $\KET{\Phi}=D^{-1/2}(\pm\phi_1\pm\phi_2\ldots\pm\phi_D)$.\cite{ECKERTone,VEDRALone}
We then calculate
$\KET{e^{-itH/2}\Phi} = \sum_{n=1} b_n(t/2)\KET{\phi_n}$
and use
\begin{equation}
\BRACKET{\Phi}{e^{-itH}\Phi}=\BRACKET{e^{itH/2}\Phi}{e^{-itH/2}\Phi}=\sum_{n=1}^D |b_n(t/2)|^2
\end{equation}
to obtain the diagonal matrix element.
Each of these steps executes very efficiently on a QC.

Remains the question how many samples $S$ are needed
to compute the energy and specific heat to high accuracy.
According to the central limit theorem the statistical error on the results
vanishes as $1/\sqrt{S}$. However, as we demonstrate below,
the application of our QA to a highly non-trivial quantum many-body
system provides strong evidence that this error also {\bf decreases} with
the system size.
For systems of 15 qubits or more, we find that
taking $S=20$ samples already gives very accurate results.
Our experimental finding that the statistical error
on $\BRACKET{\Phi}{e^{-itH}\Phi}$ for randomly chosen $\KET{\Phi}$ decreases with $D$
gives an extra boost to the efficiency of the QA.

\section{Soft Quantum Computer and Quantum Algorithm}

The method described above has been tested on our
Soft Quantum Computer (SQC). The SQC used to compute the results
presented in the present paper is a hard-coded version, derived from
of a more versatile SQC discussed elsewhere.\cite{XXXXXX}
Our SQC solves the time-dependent Schr\"odinger
equation (TDSE)

\begin{equation}
 \label{schrod}
i{\partial \over\partial t} \KET{\Psi(t)}= H \KET{\Psi(t)}
,
\end{equation}

for a quantum many-body system described by the spin-1/2 Hamiltonian
\begin{equation}
 \label{hamiltonian}
H=-\sum_{i,j=1}^L\sum_{\alpha=x,y,z} J_{i,j}^\alpha S_i^\alpha S_j^\alpha
-\sum_{i=1}^L\sum_{\alpha=x,y,z} h_{i}^\alpha S_i^\alpha
,
\end{equation}
where the first sum runs over all pairs $P$ of spins (qubits),
$S_i^\alpha$ ($\alpha=x,y,z$) denotes the $\alpha$-th component of the spin-1/2
operator representing the $i$-th spin,
$J_{i,j}^\alpha $ determines the strength of the interaction between
the spins at sites $i$ and $j$,
and $h_{i}^\alpha$ is the (local magnetic) field acting on the $i$-th spin.
The number of qubits is $L$ and $D=2^L$.
Hamiltonian~(\ref{hamiltonian}) is sufficiently
general to capture the salient features of most physical models of QC's
(our SQC also deals with time-dependent external fields).

According to~(\ref{schrod}) the QC will evolve in time
through the $D\times D$ unitary transformation $U(t)=e^{-itH}$.
We now describe the QA that computes $U(t)\KET{\Phi}$ for arbitrary $\KET{\Phi}$.
\magweg{If all eigenvalues and eigenvectors
are known then $\KET{\Psi(t+\tau)}=U(\tau)\KET{\Psi(t)}$ can be calculated
directly. However we have assumed that this data is not available so we take another
route.\cite{HDRCPR}
to set up an algorithm to solve the TDSE~(\ref{schrod}). }
Using the semi-group property of $U(t)$ to
write $U(t)=U(\tau)^m$ where $t=m\tau$, the main step is to
replace $U(\tau)$ by a symmetrized product-formula approximation.\cite{HDRCPR}
For the case at hand it is expedient to take
\begin{equation}
U(\tau)\approx {\widetilde U(\tau)} =
e^{-i\tau H_z/2}
e^{-i\tau H_y/2}
e^{-i\tau H_x}
e^{-i\tau H_y/2}
e^{-i\tau H_z/2}
,
\end{equation}
where
\begin{equation}
H_\alpha=-\sum_{i,j=1}^L J_{i,j}^\alpha S_i^\alpha S_j^\alpha
-\sum_{i=1}^L h_{i}^\alpha S_i^\alpha
\quad;\quad\alpha=x,y,z
.
\end{equation}
\magweg{Other decompositions\cite{PEDROone,SUZUKItwo}
work equally well but are somewhat less efficient.}
Evidently ${\widetilde U(\tau)}$ is unitary and hence the algorithm
to solve the TDSE is unconditionally stable.\cite{HDRCPR}
\magweg{It can be shown that
$\Vert U(\tau) - {\widetilde U(\tau)}\Vert \le c \tau^3$, implying
that the algorithm is correct to second order in the time-step $\tau$.\cite{HDRCPR}
Usually it is not difficult to choose $\tau$ so small that
for all practical purposes the results obtained can be considered
as being ``exact''.
Moreover, if necessary, ${\widetilde U(\tau)}$
can be used as a building block to construct higher-order 
algorithms.\cite{HDRone,SUZUKItwo2,HDRKRMone,SUZUKIthree}
}

As basis states $\{\KET{\phi_n}\}$ we take
the direct product of the eigenvectors of the
$S_i^z$ (i.e. spin-up $\KET{\uparrow_i}$ and spin-down $\KET{\downarrow_i}$).
In this basis, $e^{-i\tau H_z/2}$ changes the input state by altering
the phase of each of the basis vectors.
As $H_z$ is a sum of pair interactions it is trivial to rewrite this operation
as a direct product of 4x4 diagonal matrices (containing
the interaction-controlled phase shifts) and 4x4 unit matrices.
Still working in the same representation, the action of $e^{-i\tau H_y/2}$
can be written in a similar manner but the matrices that contain the
interaction-controlled phase-shift have to be replaced by
non-diagonal matrices. Although this does not present a real problem it is
more efficient and systematic to proceed as follows.
Let us denote by $X$($Y$) the rotation by $\pi/2$ of each spin
about the $x$($y$)-axis. As
\begin{equation}
 \label{last}
e^{-i\tau H_y/2}=XX^\dagger e^{-i\tau H_y/2}XX^\dagger
=X e^{-i\tau H_z^\prime/2}X^\dagger ,
\end{equation}
it is clear that the action of $e^{-i\tau H_y/2}$ can be computed by
applying to each qubit, the inverse of $X$
followed by an interaction-controlled phase-shift and $X$ itself.
The prime in~(\ref{last}) indicates that $J_{i,j}^z$ and $h_{i}^z$ in $H_z$
have to be replaced by $J_{i,j}^y$ and $h_{i}^y$ respectively.
A similar procedure is used to compute the action of
$e^{-i\tau H_x}$. \magweg{: We only have to replace $X$ by $Y$.}

Our SQC carries out ${\cal O} \left(P\,2^L\right)$ operations to perform the trans\-formation
$e^{-i\tau H_z/2}$ but a QC operates on all qubits simultaneously
and would therefore only need \ORDER{P} operations.
The operation counts for $e^{-i\tau H_x}$ (or $e^{-i\tau H_y}$)
are \ORDER{(P+2) 2^L} and \ORDER{P+2} for the SQC and QC respectively.
On a QC the total operation count per time-step is \ORDER{3\,P+4}.

\magweg{Finally we estimate how many time-steps $N$ are needed to obtain
$E$ and $C$ with acceptable accuracy.
First we note that the number of read-outs of the QC can be minimized
by using the Nyquist sampling theorem and
the fact that $H$ is bounded from above and below.
Indeed, it is sufficient to measure at intervals
$\Delta t=\pi/E_{max}$ where $E_{max}=\max_i |E_i|$.
If $\tau_0$ denotes the maximum value of $\tau$
that yields an accurate approximation ${\widetilde U(\tau)}$
we set $\tau=\min\{\tau_0,\Delta t\}$ and solve the TDSE for $N$ time steps.
$N$ determines the resolution $\Delta \epsilon=\pi/N\tau$
of the discrete Fourier transform (note that $E_{max}$ increases
linearly with $L$ whereas the number of eigenvalues in the
interval $[-E_{max},E_{max}]$ increases exponentially with $L$).
Taking a very pessimistic point of view we may expect to obtain
accurate values for $E$ and $C$ as long as $T>\Delta\epsilon$.
It is important to note that increasing this accuracy
by a factor of two only requires running the QC twice as long: The
error decreases linearly with computation time.
}
\section{Application}
The QA described above has been tested on our SQC
by simulating the antiferromagnetic spin 1/2 Heisenberg model 
with $J=-1$
\magweg{(i.e. $J_{i,j}^\alpha =-1$ if $i$ and $j$ are nearest neighbors and
$J_{i,j}^\alpha =0$ otherwise, $h_i^\alpha=0$)}
on triangular lattices of $L=6,10,15,21$ sites, subject to free boundary conditions.
The ground-state properties of this model can be computed by standard
sparse-matrix techniques, see e.g. Ref.~\citen{DEUTCHER}.
The low temperature properties of this model are difficult
to compute by conventional Quantum Monte Carlo (QMC) 
methods.\cite{HATANOone} 
The presence of frustrated interactions leads to the sign problem\cite{HATANOone}
that is very often encountered in QMC 
work.\cite{HDRQMC,HDRSD}

\begin{figure}[t]
\parbox{\halftext}{
\epsfxsize=6.0cm
\begin{center}\epsffile{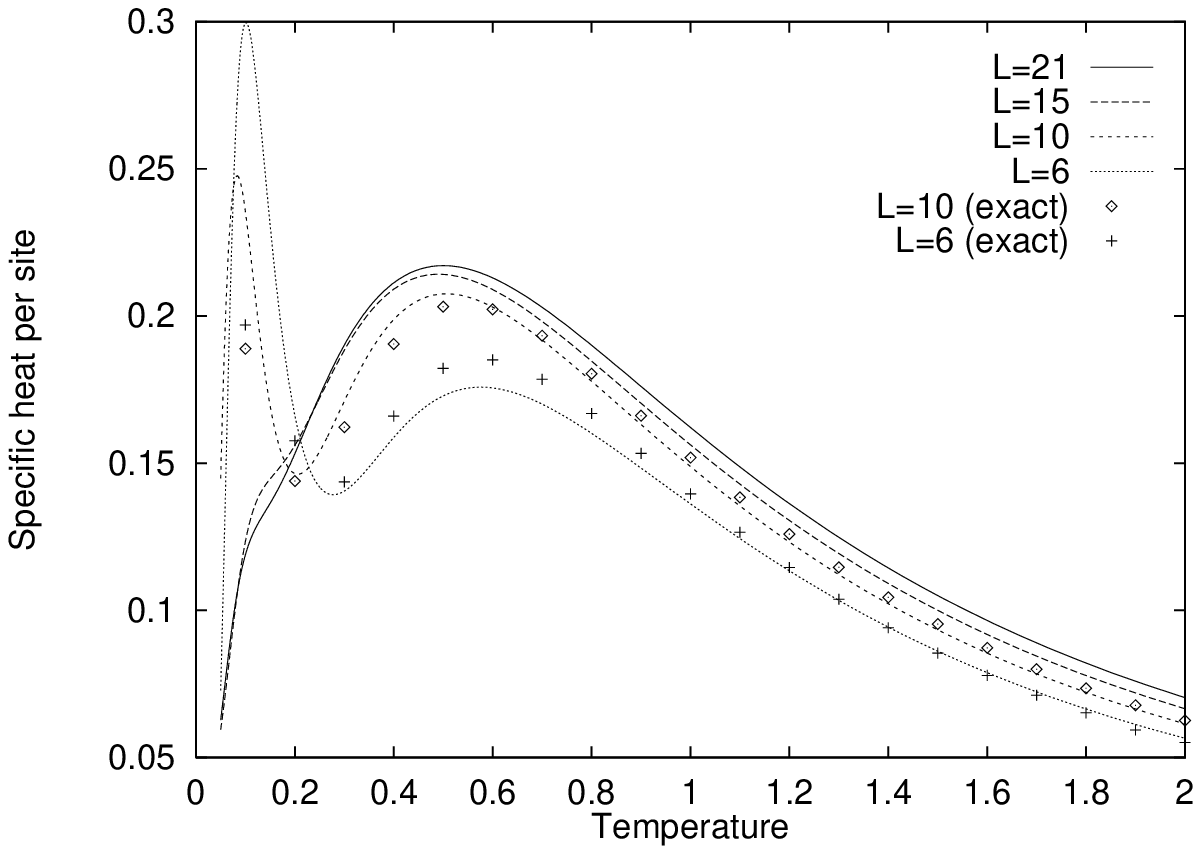}\end{center}
\caption{ Specific heat per site as a function of the temperature.}
\label{fig:specheat}}
\hspace{4mm}
\parbox{\halftext}{
\epsfxsize=6.0cm
\begin{center}\epsffile{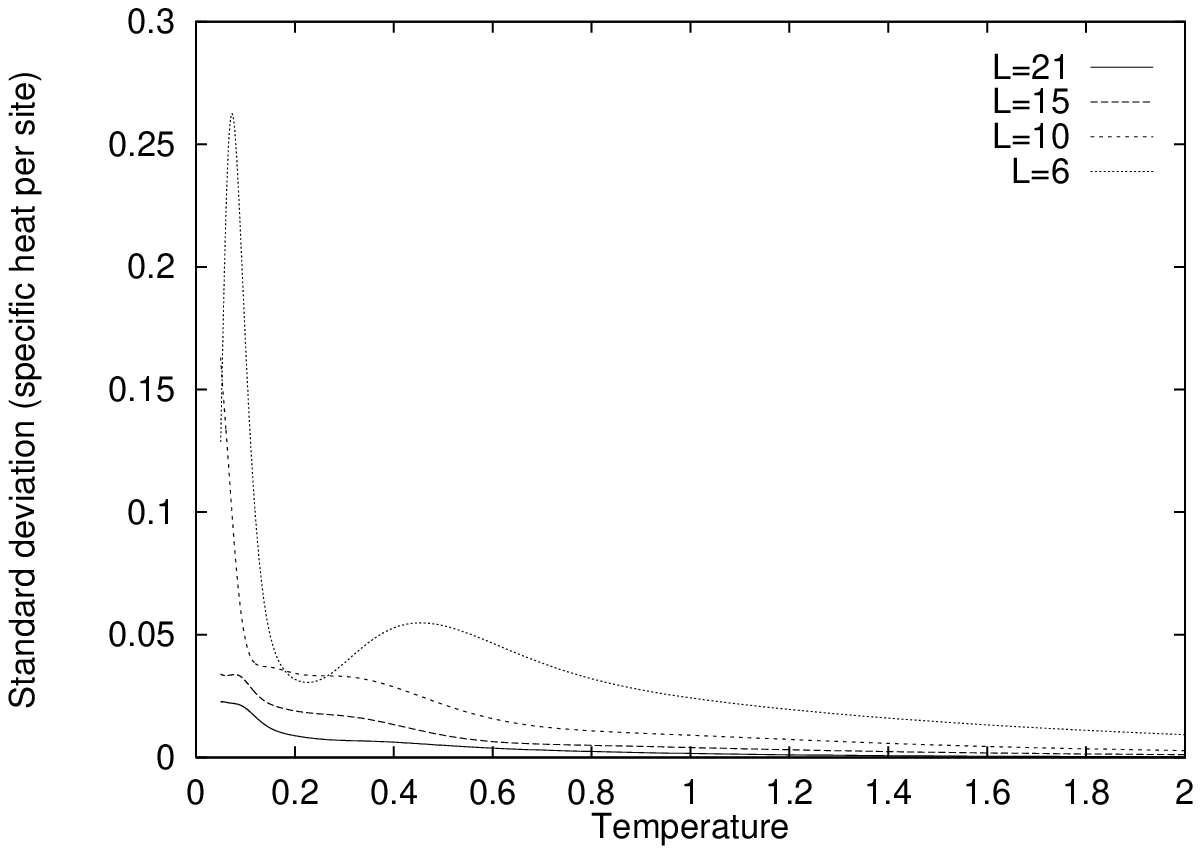}\end{center} 
\caption{Standard deviation on the specific heat per site.}
\label{fig:stddev}}
\end{figure}

In Fig.~\ref{fig:specheat} we present some SQC results for the specific heat per site $E/L$ as a function
of the temperature. The number of samples $S=20$ in all cases.
Also shown is data obtained by exact diagonalization
of $H$ for $L=6,10$.\footnote{The calculation of all 32768 (2097152) eigenvalues of the $L=15$ ($L=21$) system
by standard linear algebra methods requires tremendous computational resources}
The SQC and exact results differ significantly
for temperatures where the specific heat exhibits a sharp peak.
This is related to the presence of a gap in the low-energy part of the spectrum
and the random choice of the $\KET{\Phi}$'s. In this low-temperature
regime where only a few of the lowest eigenvalues contribute,
random fluctuations can have a large effect.
However this is not really a problem: Knowing the DOS
it is not difficult to determine the precise values of these few eigenvalues
and compute $C$ directly, without invoking~(\ref{min8}).
\magweg{We do not present such an analysis here because the emphasis of this
paper is on the QA, not on advanced methods of processing DOS data.}
In Fig.~\ref{fig:stddev} we show results for the standard deviation (SD) on $C/L$,
calculated from the same data\magweg{ set of $S=20$ samples}.
The most remarkable feature of the SD is its dependence on the
system size: The larger the system, the smaller the SD.
Unfortunately we cannot yet
offer a theoretical basis for this observed decrease.
At very low temperature the SD on $C/L$ goes to zero because
only one eigenstate (i.e. the ground state) effectively contributes.
The large values of the $L=6,10$ low temperature SD data
reflect the fact that in this regime the SQC and exact results differ considerably
and indicate that for these system sizes more than $S=20$ samples are necessary to
obtain accurate results. On the other hand, except for test purposes,
we wouldn't use a QC to simulate a 10-site system because it can easily
be solved exactly on an ordinary computer.

\magweg{The decrease of the SD with system size gives an extra boost to the
efficiency of our QA. 
Obviously for a non-interacting system, a randomly chosen initial state
will lead to the correct answer, with a SD that decreases with the
system size. Although our results strongly suggest that a similar phenomenon
is also at work in the case of an interacting system, we believe that more work is
necessary to establish, and perhaps prove, that this is a general
feature of the QA discussed above.}

\section{Summary}
We have described a quantum algorithm to determine
the distribution of eigenvalues of a quantum many-body system
in polynomial time. From these data thermal equilibrium
properties of the system can be computed directly.
The approach has been illustrated by numerical calculations
on a software emulator of a physical model of a quantum computer.
Excellent results have been obtained, suggesting
a new route for simulating experiments on quantum systems on a quantum computer.
However, implicit in the formulation of this physical model of the quantum computer
is the assumption that each physical spin represents one qubit.
If this were not the case, the quantum computer will operate with much less efficiency.

\section*{Acknowledgements}
Support from the Dutch ``Stichting Nationale Computer
Faciliteiten (NCF)'' and from the Grant-in-Aid for Research from the
Japanese Ministry of Education, Science and Culture is gratefully
acknowledged.


\end{document}